\documentclass[11pt]{article}
\usepackage[dvipsnames]{xcolor}

\usepackage{url,color,tikz,dblfloatfix}
\usepackage{mathrsfs}
\usepackage{pgfplots}
\usepackage{graphicx}
\usepackage{subcaption}
\usepackage{bbm}
\usepgfplotslibrary{groupplots}
\pgfplotsset{
    compat=1.8,
    filter discard warning = false }
\usepackage{tikzscale}

\usepackage[english]{babel}
\usepackage[utf8]{inputenc}

\usepackage{amsmath,amsthm,amssymb,bbm}
\usepackage{graphicx}
\usepackage[T1]{fontenc}
\usepackage[a4paper,bindingoffset=0.5cm,left=2cm,right=2cm,top=2cm,bottom=2cm,footskip=.8cm]{geometry}
\usepackage[md]{titlesec}
\usepackage{color}

\usepackage{mathrsfs}
\usepackage[mathscr]{euscript}

\usepackage{mathrsfs}

\definecolor{light}{gray}{0.8}

\usepackage[english]{babel}
\usepackage{amsmath,amsthm,amssymb,bbm,cancel}
\usepackage{graphicx}
\usepackage[a4paper,bindingoffset=0.5cm,left=2cm,right=2cm,top=2cm,bottom=2cm,footskip=.8cm]{geometry}
\usepackage{color,tikz}

\newtheorem{proposition}{Proposition}[section]

\newcommand{\I}{\mathrm{I}}

\newcommand{\E}{\mathbb E}

\usepackage{bm}

\theoremstyle{remark}

\newcommand{\lo}{{\rm o}}
\newcommand{\di}{{\rm d}}

\def\bO{\boldsymbol 1}
\def\bP{\boldsymbol \pi}

\begin{document}
\title{Transient Provisioning and Performance Evaluation for Cloud Computing Platforms: A Capacity Value Approach}
\author{Brendan~Patch\thanks{Corresponding author. Email: b.patch@uq.edu.au}~\thanks{The School of Mathematics and Physics, The University of Queensland, Qld 4169, Australia.}~\thanks{Korteweg-de Vries Institute for Mathematics, University of Amsterdam, 1090 GE  Amsterdam, The Netherlands.} \quad and\quad Thomas Taimre\footnotemark[2].}

\maketitle

\begin{abstract}
User demand on the computational resources of cloud computing platforms varies over time.
These variations in demand can be predictable or unpredictable, resulting in `bursty' fluctuations in demand.
Furthermore, demand can arrive in batches, and users whose demands are not met can be impatient.
We demonstrate how to compute the expected revenue loss over a finite time horizon in the presence of all these model characteristics through the use of matrix analytic methods.
We then illustrate how to use this knowledge to make frequent short term provisioning decisions --- transient provisioning.
It is seen that taking each of the characteristics of fluctuating user demand (predictable, unpredictable, batchy) into account
can result in a substantial reduction of losses.
Moreover, our transient provisioning framework allows for a wide variety of system behaviors to be modeled and gives simple expressions for expected revenue loss
which are straightforward to evaluate numerically.
\end{abstract}


\section{Introduction}
Highly complex systems are becoming an integral contributor to the productivity of many industries. The introduction of these systems is accompanied by an increase in the demand for computational resources. Distributed cloud computing platforms have emerged as the leading method for provision of these resources to end users. In addition to the environments available online (e.g.\ Amazon EC2, Microsoft Azure, Google AppEngine, GoGrid), many private organizations and universities now have computer clusters that allow users to distribute computing tasks across many nodes. This substantially reduces the need for each user to have expensive individually held computing resources that become idle when not needed or which are impractical to store at the users' geographical location.

Distributed computing constitutes a substantial portion of the energy consumption in modern computer and communication networks \cite{Mann2015}.
As such, well designed provisioning policies, which match the availability of resources with the demand for resources remains an active area of research \cite{Mann2015}. An obvious avenue to reducing the energy use of a  distributed cloud platform is to switch compute nodes off, or place them into a power saving mode, when they are not needed. For example, in \cite{Zhang2012} it is estimated that perfectly provisioning capacity to match demand in a production compute cluster at Google would result in a 17--22\% reduction in energy use.

Resource allocation problems of this type naturally fall into the realm of queueing theory.
Specifically, the {\em loss network model} has been extensively used to analyze circuit switched systems in which tasks arrive randomly throughout time, require a random service time, and are lost if the resources required to begin their service are not available at the time of their arrival (see for example \cite{Kelly1991,QueueBook}).
In these models, the key quantity of system capacity is usually viewed as static, unable to be altered in response to short term fluctuations in system demand.
As such, work on capacity selection is typically based on equilibrium properties of the system.
Probably the most famous result of this type is Erlang's \cite{Erlang1918} expression for the probability that a task arriving to the system in steady state is blocked from entry, when tasks arrive according to a Poisson process with rate $\lambda$, have a mean service time of $\mu^{-1}$, and there are $m$ servers (an $M/G/m/m$ queueing system):
\begin{equation}\label{eq:Erlang}
\frac{(\lambda/\mu)^m/m!}{\sum_{i=0}^m (\lambda/\mu)^i/i!}\,.
\end{equation}

Much of the literature on the analysis of cloud computing platforms, which we review in Section~\ref{sec:RelatedWork}, develops performance measures (e.g.\ probability of a blocked task, waiting times, response times) from an equilibrium perspective. In order to improve the matching between provisioned capacity and demand throughout time, and to achieve the energy savings which motivate our work, it is necessary to take a transient view of the system. Due to the elasticity, or short term capacity flexibility, of distributed cloud computing platforms this is especially relevant in our case \cite{Mann2015}.

Moving away from equilibrium analysis of these systems in favor of transient analysis allows provisioning to be performed over short time intervals in a way that is dependent on the current state of the system and knowledge of the arrival rate over the near future. Since obtaining analytic results for the transient distribution of loss network type systems is notoriously difficult, one often resorts to numerical inversion of Laplace transforms \cite{Abate1998,Knessl2015} or approximations \cite{Mandjes2001}.  In \cite{Chiera2002}, its companion \cite{Chiera2005}, and more recently \cite{Braunsteins2016} and \cite{Patch2015}, a useful alternative to the consideration of the transient distribution for queueing type models is proposed.
In these papers the authors assume that tasks which fail to enter a loss system due to capacity constraints result in the system's manager incurring a predetermined amount of lost revenue.
By comparing the amount of lost revenue during a finite time interval $[0,\, t]$ that results from different capacity choices, the authors are able to determine buying and selling prices for a unit of capacity using only information on the value of lost tasks and the current number of tasks in the system.
They call the function underlying these rules the \emph{capacity value function}.

The results in \cite{Chiera2005} and \cite{Chiera2002} are, however, derived using delicate manipulations of orthogonal polynomials, and it is difficult to use the same analytical techniques to generalize these findings to models which are more applicable to the distributed cloud computing setting.
In this paper we overcome this issue by making the pivotal observation that the capacity value function can be expressed in terms of matrix inverses and exponentials using \emph{matrix analytic methods} (MAMs).
Moreover, by adding a term to the capacity value function that reflects the operating costs of maintaining different levels of capacity over time we allow the trade-off between energy use and service degradation, that is controlled through the provisioning decision, to be explicitly modeled.

Based on these observations, in this paper we show how to utilize the well established MAM literature to effectively obtain a transient performance measure, similar to the capacity value function, for a wide variety of potential cloud computing models. The extension of the capacity value function methodology to more general settings is the main contribution of this paper. Although our framework is widely applicable, for clarity we illustrate it using a model that incorporates the following features:
\begin{itemize}
\item Batch jobs.
\item Bursty arrival process (predictable and unpredictable).
\item A buffer.
\item Abandonments due to impatience.
\end{itemize}
In the outlook we provide a more detailed discussion on the types of settings we envisage our framework is applicable to. In general we are concerned with any cloud system where tasks are sent to the system by users with the aim of undergoing processing before eventual departure. We envisage that in order to be processed the tasks may require access to a database or directory, specific software (e.g.\ a compiler), and/or specific hardware (e.g.\ CPU cores or RAM). Throughout the paper we have emphasized the use of our framework on cloud platforms, there is nothing however in principle that stops it from being applied to cloud infrastructure or applications.

In our model tasks arrive to the system according to a \emph{batch Markovian arrival process} (BMAP). We call an arrival a job and each server request that a job makes is a task.
Specifically, allowing batch arrivals means that jobs may request a random number of units of server upon arrival.  When the random variable governing the number of tasks that may be requested by a job has higher expectation or takes on a greater range of values we consider the arrival process to be more `batchy'.  It is important to highlight that allowing tasks to arrive in batches can be viewed as modeling jobs as Erlang distributed with a random shape parameter, which is substantially more general than the exponential distribution employed in \cite{Chiera2005} and \cite{Chiera2002}. Our notion of a batch is similar to the notion of a `supertask' in \cite{Khazaei2013b}.

In addition, this arrival process allows time varying behavior to be modeled.  We incorporate `burstiness' (or unpredictability) into our model by using the property of BMAPs that the arrival rate of tasks for these processes may change randomly throughout time according to an underlying modulating Markov process. For simplicity we suppose that this underlying process alternates between a baseline state and a state where the arrival rate is increased.  We view the difference between the baseline arrival rate and the randomly increased arrival rate, as well the frequency with which the randomly increased rate is expected to occur, as measures of the system's burstiness. For example, if the system experiences large, frequent, unpredictable increases in the arrival rate then we would say that the system is more bursty than a system experiencing infrequent minor increases in the arrival rate. It will become clear to the reader how this simplifying assumption can easily be relaxed to permit a modulating Markov process with any finite number of states, rather than just two (and we provide details on this in the concluding section). In addition, our framework can be applied in the case that it is known when a change in the arrival rate will occur during our planning horizon. We call instances of known changes in the arrival rate predictable bursts.

The presence of a buffer allows a cloud platform provider to store tasks which arrive when all of the servers are in use, so that the tasks may be processed once a server becomes available. In many cases it may be possible for a user to be aware that their task is waiting in the buffer, rather than actively undergoing service. In a competitive environment the user may choose in such a circumstance to attempt service with an alternative cloud platform provider (i.e.\ abandon the system). Accounting for user behavior such as this may be highly beneficial to cloud providers, and a key advantage of our framework is that such extensions are often easily incorporated.

Our key result is an explicit matrix expression for the expected lost revenue during $[0,\,t]$ when $m$ servers are active, tasks may wait in a buffer of size $r$, there is a cost per unit time per unit of server, and a potentially different cost per unit time per unit of buffer.
In addition to the revenue lost from a task failing to enter the system we also suppose that when tasks which are waiting for service in the buffer abandon the system a loss is also incurred by the system manager. We suppose that tasks will wait an exponential amount of time before abandoning.

\begin{figure}[h]\centering
\includegraphics[scale=1.4]{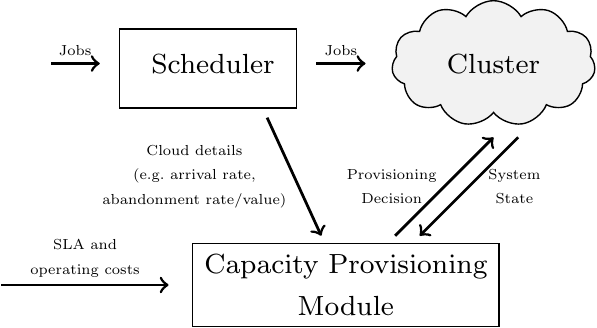}
\caption{Interaction of our framework with a cluster and existing cloud platform architecture.}
\label{fig:IdealisedScenario}
\end{figure}

In the context of the model just described we show how our transient performance measure (expected lost revenue during $[0,\,t]$) can be used by system managers to make short term provisioning decisions.
Furthermore, analyzing the performance of the system in terms of revenue losses due to blockages and abandonments is particularly relevant in this setting since it realistically reflects the penalties imposed on service providers associated with violations of service level agreements.
Our framework provides a key prerequisite for the development of a capacity provisioning module that could be combined with other tools in a realistic cloud setting to provide improvements in performance. In Fig.~\ref{fig:IdealisedScenario} we illustrate how our framework would combine with existing cloud architecture such as micro services, service composition, and load balancing (summarized as a scheduler for simplicity) and a cloud computing platform (represented as a cluster). In this setting, these other aspects of the cloud architecture will have implications for the arrival rate of jobs to a particular cloud computing platform, upon taking this into account our framework provides a service that is complementary to the existing architecture.

Any improvements in performance from using our framework to provision a cloud computing platform will clearly depend on the parameterization of the real world system to which it is applied.
Parameterization of MAPs is an area of ongoing research (see e.g.\ \cite{Breuer2002,Buchholz2010,Buchholz2009,Buchholz2017,Casale2008}), and tools are becoming readily available for real world managers to utilize. Finally, some cloud providers now allow users to set automatic decision rules on when to scale up or down capacity (e.g.\ \cite{AmazonSite}) --- after parameterization of an appropriate application specific model our framework could be applied to guide these decisions.

The analytical expressions that we detail and the increment in performance that is gained when using them in place of traditional equilibrium based performance measures is illustrated through several examples.
In these examples we see that when a system is subject to regular predictable bursts our method can lead to a substantial reduction in losses.  More modest improvements in performance are also seen when fluctuations in demand are unpredictable.

An important aspect of our framework is that the computations needed for its online implementation do not {\em necessarily} need to be performed online (although in many cases an online implementation would be possible); the manager only needs access to the decision corresponding to each potential system state. For a system with $m$ servers, $r$ buffers, and simple burst behavior, obtaining these decision rules will require the inverse and matrix exponential of matrices with dimension $2(m{+}r{+}1)$ to be found.  Using the basic form of Gauss--Jordan elimination, a matrix inverse can be computed in ${\text O}(n^3)$ steps. Various algorithms, each with their individual strengths and weaknesses, are available to compute the matrix exponential (c.f.\ \cite{Sidje1998}): for a discussion of these issues see \cite{Moler1978}. For large systems it may be more practical to store decisions (i.e.\ two $2(m+r+1)$-dimensional vectors which give the optimal server and buffer choices for each system state) and access them when needed.

\subsection{Related Work}\label{sec:RelatedWork}
One of the main triggers behind our work is the observation that much of the work on performance evaluation of cloud platforms is based on a description of the system in equilibrium. Our work is based on a {\em transient} description of the system, which is prudent since: i) the system may never reach equilibrium between changes in demand, and ii) a transient procedure allows the manager to take advantage of information on the present state of the system when making a provisioning decision.

In \cite{Khazaei2012} Khazaei et al.\ propose an $M/G/m/m{+}r$ queueing system as an approximation to the type of real world distributed cloud computing platform system we are interested in.
In their model compute tasks arrive to an $m$ server system according to a homogeneous Poisson process, have a generally distributed service time, and are able to occupy $r$ input buffer places if the system is already processing $m$ tasks upon arrival, but are otherwise lost.
Through an analysis based on the equilibrium state of the system the authors are able to determine the relationship between the number of servers and input buffer size and equilibrium performance indicators such as mean queue size, blocking probability, and the probability that a task will enter service immediately upon arrival.
More recently Atmaca et al.\ \cite{Atmaca2016} have provided a generalization of this work that uses Phase-type distributions to model service times and the time between arrivals. There are also generalizations of the model so that bursty (see e.g.\ \cite{Zhang2016}) and batchy (see e.g.\ \cite{Khazaei2011,Khazaei2013,Khazaei2013b}) behavior can be investigated, again from an equilibrium perspective.

A related presentation is given in \cite{Tan2015} where Tan and Xia model the distributed cloud from a revenue management perspective as a multi-class loss network with jobs of different types arriving according to a general renewal process.
While in \cite{Bruneo2014} Bruneo gives a model based on stochastic reward nets that is scalable to very large system sizes and is flexible enough to be adapted to different scenarios --- similar performance metrics are again analyzed from an equilibrium point of view.

In \cite{Maccio2015} Maccio and Down introduce a model that has a single server which switches between on and off states according to the length of the queue. Again using steady state analysis, several similar performance metrics are connected to the system parameters and some key observations on how the system behaves are made.

A notable exception to the dominant equilibrium analysis is a discrete time model predictive control based approach given by Zhang et al.\ in \cite{Zhang2012}. The empirical approach taken in their work is presented as a promising initial step towards provisioning cloud computing platforms and represents an approach that is methodologically complementary to the one we present. Also of note is the recent work of Leeuwaarden et al.\ in \cite{Leeuwaarden2016}, where a cloud provisioning algorithm based on heavy traffic approximations and asymptotic analysis is introduced and verified using simulation.

Although MAMs have previously been used to study cloud computing platforms (see e.g.\ \cite[Chapter 21]{bookHarcholBalter2013}), they have not yet been used to evaluate effective \emph{transient} performance measures. A related modeling formalism that also shows promise for the effective transient analysis of cloud computing platforms is stochastic petri nets. A key piece of work investigating this avenue is \cite{Ghosh2010}, where a similar notion of transient and equilibrium losses from blockages is investigated. The authors of this work do not incorporate system operating costs into their performance measure, and so the trade off between alternative capacity choices and blocking is not explicitly considered. Rather the focus is on determining the resiliency of the system to exogenous uncontrolled changes in capacity (for example due to system failure).

In the next section we will detail our model of a cloud computing platform, and then in the subsequent section we will develop an encompassing model that takes the system model as an input to allow our performance measure to be computed. This is followed by some examples that show how to use the performance measures to make short term provisioning decisions.

\subsection{Organization}
The remainder of this paper is structured as follows. In Section~\ref{sec:model} we give a formal description of our illustrative model for a cloud computing platform.
Section~\ref{sec:PEmodel} develops a framework around this model that allows us to present a method of performance evaluation in Section~\ref{sec:performance}.
In Section~\ref{sec:examples} we illustrate our method.
We provide an outlook to future research in Section~\ref{sec:conclude}.

\section{Model of Cloud Computing Platforms}\label{sec:model}
In this section we introduce a model of cloud computing platforms that reflects the features we discussed in the introduction and which we will later use in the development of our novel encompassing performance evaluation model.
We have been careful to be very clear about the assumptions of the model which we use to illustrate the framework by detailing the features that we {\em do} include, while simultaneously emphasizing the scope of applicability by providing some detail on the features that we {\em do not} include.

We assume that each arrival to the system consists of at most $\ell$ tasks, each of which requires its own server (e.g.\ a CPU core) to be processed or unit of buffer to be held in.
Furthermore, we assume that the system exists in a random environment where traffic usually arrives according to some `normal' rate, but occasionally arrives at some other `bursty' rate.
Throughout this paper we use the term \emph{rate} in the following sense: when an event occurs at rate $\lambda$ at time $t$ we take that to mean that the probability of the event occurring during $[t, t+h]$ is $\lambda h + \lo(h)$ where $\lo(h)/h \to 0$ as $h\to 0$, i.e.\ $\lo(h)$ converges to zero faster than a linear function of $h$.
Let $Y(t) \in \{1,2\}$ equal 1 when arrivals are occurring at the normal rate at time $t$ and equal 2 when arrivals are occurring at a bursty rate at time $t$.
Moreover, when $Y$ is in state 1 it transitions to state 2 at rate $\alpha$, and when $Y$ is in state 2 it transitions to state 1 at rate $\beta$.
Let the number of tasks that a job brings to the system be governed by the random variable $K$, we assume that $K$ has finite support. We denote the arrival rate of jobs consisting of $k \in \{1,\dots, \ell\}$ tasks when $Y(\tau)=y$ at time $\tau>0$ by $\lambda_y^{(k)}$.
Note that the transitions of $Y$ govern the frequency and duration of unpredictable bursty periods.
It is a simple matter to extend our model to have multiple burst types of different frequency and duration.

Let $X(t) \in \{0, 1, \dots, m, m+1, \dots, m+r\}$ be the number of tasks being processed at time $t$ by a cloud platform with the preceding arrival process, $m$ servers, and a buffer of size $r$.
When $X(t)\le m$, then all of the tasks in the system are being served, however when $X(t) > m$, then $m$ tasks are being served and $X(t)-m$ are waiting in the buffer.
We assume that tasks require an exponentially distributed service time with mean $\mu_s^{-1}$. Therefore, the set of parameters $(\lambda_y^{(k)}, y \in \{1,2\}, k \in \{1,\dots,\ell\})$ and $\mu_s^{-1}$ together allow for the distribution of job processing times to be Erlang distributed with a random shape parameter $K$ and rate parameter $\mu_s^{-1}$.

\begin{figure*}[h]\centering
\includegraphics{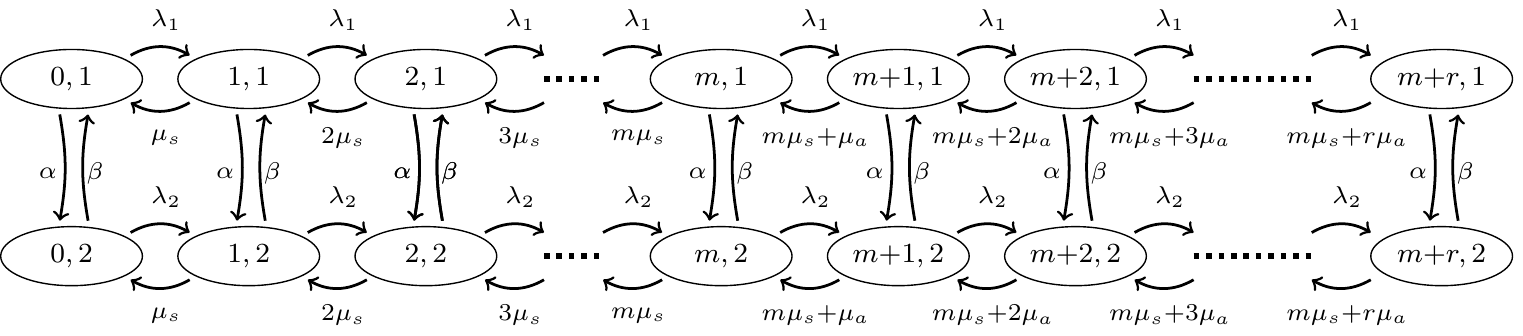}
\caption{State transition diagram for a distributed computing platform with time homogeneous unpredictable arrivals, $m$ servers, $r$ units of buffer, and arrivals always consisting of a single task ($\ell = 1$).}
\label{fig:states}
\end{figure*}

If a job consisting of $k$ tasks arrives to the system when there are fewer than $k$ servers or buffer units available, (i.e.\ $X(t)$ greater than $m+r-k$), then the task is blocked from entry and lost.
Furthermore, we also incorporate impatience into our model.
When a task is in the buffer, it will wait up to an exponentially distributed amount of time with mean $\mu_a^{-1}$ for service to commence, but will otherwise abandon the system without being served.
The states and transitions of this finite state Markov process are illustrated in Fig.~\ref{fig:states} for the case when each arrival to the system always consists of only a single task.
For clarity, we also compactly summarize the full set of transitions in Table~\ref{tab:rates}.

At this point we reiterate that the performance evaluation framework we develop in the next section can also be applied to many models other than the one we have detailed in this section. For example, the servers in a cloud may not necessarily be homogeneous (as we have assumed). So long as the simple task-based workload that we assume applies to the system under consideration, the processing rates of each server can be modified through the entries in Table~\ref{tab:rates}. To see this, consider the case that a system possesses a fixed amount of processing power that is shared between the tasks being processed, then $x\mu_s$ would be replaced by $\mu_s/x$ in the table.

\begin{table*}[h]
\caption{Transition rates of our bursty batch cloud computing platform.}
\begin{center}
\begin{tabular}{l |c | c}\label{tab:rates}
\hspace{7mm}Transition & Rate at time $\tau$ & States\\
\hline \hline
$(x,\, 1) \rightarrow (x,\, 2)$ & $\alpha$ & $0 \le x \le m+r$\\
$(x,\,2) \rightarrow (x,\,1)$ & $\beta$ & $0 \le x \le m+r$\\
$(x,\, 1) \rightarrow (x+k,\, 1)$ & $\lambda_1^{(k)}$ & $0 \le x \le m+r-k$, $1 \le k \le \ell$\\
$(x,\, 2) \rightarrow (x+k,\, 2)$ & $\lambda_2^{(k)}$ & $0 \le x \le m+r-k$, $1 \le k \le \ell$\\
$(x,\,y) \rightarrow (x-1,\, y)$ & $x\,\mu_s$ & $ 0 < x \le m$, $y \in \{1,2\}$\\
$(x,\,y) \rightarrow (x-1,\, y)$ & $m\,\mu_s+(x-m)\,\mu_a$ & $ m < x \le m+r$, $y \in \{1,2\}$\\
\hline \hline
\end{tabular}
\end{center}
\end{table*}

\section{Encompassing Performance Evaluation Model}\label{sec:PEmodel}
Now that we have a general model of how the system operates, we must define some additional stochastic processes that allow us to perform transient performance evaluation.
To see why it is necessary to utilize these additional processes, observe that $X(t)$ does not provide any information on blocked and abandoned tasks during $[0,\,t]$ --- we must develop an encompassing model that also records these losses.
As Chiera et al.\ do in \cite{Chiera2002} and \cite{Chiera2005}, we assume that each blocked task costs the manager of the system some pre-determined amount $\theta_b > 0$.
Similarly, each task that abandons the system without being served incurs a cost of $\theta_a$.
Furthermore, we suppose that each active server results in a cost of $\theta_s$ per unit of time, and similarly each active unit of buffer results in a cost of $\theta_u$ per unit of time.
Specifically, over any time interval $[t_1,\,t_2]$ the system manager incurs a deterministic cost of $(m\theta_s+r\theta_u)\,(t_2-t_1)$ to maintain $m$ servers and $r$ units of buffer.
Similarly, when $X$ is in a state $x$ such that $x > m+r-\ell$ during a time interval $[t_1,\,t_2]$ and $Y$ is in state $y$, a loss of
\[
(t_2-t_1)\,\theta_b\,\sum_{k=m+r-x+1}^{\ell} k\,\lambda_y^{(k)}
\]
is expected to be incurred from blocked tasks.

Table~\ref{tab:costs} summarizes the notation used for different types of losses for the reader's convenience.

\begin{table}[h]
\caption{Cost parameters.}
\begin{center}
\begin{tabular}{l | l }\label{tab:costs}
Parameter & Definition \\
\hline \hline
$\theta_b$ & Lost revenue from blocked task.\\
$\theta_a$ & Lost revenue from abandoning task.\\
$\theta_s$ & Server cost per unit time. \\
$\theta_u$ & Buffer cost per unit time.\\
\hline \hline
\end{tabular}
\end{center}
\end{table}

Let $R^{m,r}_{x,y}(t)$ denote the revenue lost during $[0,\,t]$ from blocked tasks (i.e.\ tasks that attempt to enter the system when it is at capacity) when $(X(0),Y(0)) = (x,y)$ and during $[0,\,t]$ there are $m$ servers available with $r$ units of buffer. Similarly, let $A^{m,r}_{x,y}(t)$ denote the revenue lost during $[0,\,t]$ from abandonments (i.e.\ tasks that leave the buffer due to impatience) and let $M^{m,r}_{x,y}(t)$ denote the cost of operating the system for $t$ time units, each with capacity consisting of $m$ servers and $r$ units of buffer, and initial condition $(x,\,y)$. Therefore the expected revenue loss during $[0,\,t]$, under the specified initial condition $(x,\,y)$ and system size $(m,\,r)$, can be written as
\begin{align}
g_{x,y}^{m,r}(t) &:= \E [R^{m,r}_{x,y}(t)]+\E [A^{m,r}_{x,y}(t)]+\E [M^{m,r}_{x,y}(t)]\,.\label{eq:LossStandard}
\end{align}

The function $g_{x,y}^{m,r}$ is reminiscent of the capacity value function of \cite{Chiera2005} and \cite{Chiera2002}, and so we will also refer to it by that name.
For a system with unit arrivals, no bursts in the arrival process, no buffer, no abandonments, and no capacity or buffer costs $g_{x,y}^{m,r}$ reduces to the original capacity value function formulation.
Our novel augmentation of the original capacity value function with the additional operating cost term $\E [M^{m,r}_{x,y}(t)]$ is necessary for the method to be implemented as a provisioning framework.
Despite this, the key technical challenge in the evaluation of $g_{x,y}^{m,r}(t)$ lies in the computation of $\E [R^{m,r}_{x,y}(t)]$ and $\E [A^{m,r}_{x,y}(t)]$.
The foremost reason that evaluation of $\E [M^{m,r}_{x,y}(t)]$ does not present a technical challenge is that we assume system operating costs depend deterministically on $m$ and $r$.
Within this assumption, a variety of cost formulations are possible. For example, economies of scale may exist in the provisioning of servers. So that issues such as these do not distract from our primary technical contribution (which is the determination of $\E [R^{m,r}_{x,y}(t)]+\E [A^{m,r}_{x,y}(t)]$) we assume that the expected system operating costs accrue according to the following simple linear function
\begin{align*}
\E [M^{m,r}_{x,y}(t)] = (m\theta_s+r\theta_u)\,t\,,
\end{align*}
independently of $X(0)$ and $Y(0)$. In the case that operating costs were to vary deterministically with time, this function could be modified accordingly.

It is instructive to write an integral expression for $R_{x,y}^{m,r}(t)$ so that some intuition for our performance evaluation model can be obtained, with $\I\{B\}$ the indicator function for event $B$, as follows:
\begin{equation*}
\int_0^t \theta_b\left(\sum_{y=1}^2\sum_{s=0}^{\ell-1} \sum_{j=s+1}^\ell \lambda_y^{(j)}j\I{\{X(\tau) = m+r-s\}}\I{\{Y(\tau) = y\}}\right)\di \tau\,.
\end{equation*}
This random variable can be understood as the value of an accumulation of the arrivals from underlying Poisson processes that are switched `on' and `off' as needed by the pair of binary valued random processes that indicate when $X$ and $Y$ are in states that result in blockages.
The term $\E [R^{m,r}_{x,y}(t)]$ is the expected value of this integral conditioned on $X(0)=x$ and $Y(0) = y$. Soon we will see that MAMs provide a powerful and convenient avenue to evaluation of this expression.
Since abandonment losses result from transitions of $(X,Y)$ rather than holding times of $(X,Y)$, it is not straightforward to write a similar expression for $\E [A^{m,r}_{x,y}(t)]$; nonetheless MAMs may still be used for its evaluation.

The value over the planning horizon of $[0,\,t]$ of a change in $m$ and $r$ to $\tilde m$ and $\tilde r$ is
\begin{align}
g^{\tilde m, \tilde r}_{x,y}(t) - g^{m, r}_{x,y}(t)\,.\label{control1}
\end{align}
This expression is the basis of our transient provisioning framework.
By choosing values of $\tilde m$ and $\tilde r$ that maximize revenue during the chosen planning horizon, the system manager is able to improve performance.
Server and buffer space will only be active if it is expected to generate more revenue than the operating costs of having it active.
Importantly, finding these optimal values is a simple numerical procedure that only needs to be performed once for any given set of parameters.
Although not necessarily required, for large systems it may be sensible that the decision rules  be stored in memory that is fast to access.

We will now outline a novel method (that generalizes results in \cite{Braunsteins2016}, \cite{Chiera2005}, and \cite{Chiera2002}) for obtaining explicit values of \eqref{eq:LossStandard} (and therefore \eqref{control1}).
A key observation of this paper is that the process $(X(t), Y(t) : t\ge0)$, or simply $(X,\,Y)$, which gives the current number of tasks held by the system and the current mode of arrivals, can be viewed as the background process of a pair of batch Markovian arrival processes (BMAPs).
The first of the BMAPs records the number of tasks which are blocked from entry to the system during $[0,\,t]$ due to capacity constraints, while the second BMAP records the number of abandonments during $[0,\,t]$.
We will refer to these as the `blocking BMAP' and `abandonment BMAP' respectively.
The rest of this section aims to show how viewing the system in this way allows MAMs to be exploited so that the explicit computation of $\E [R^{m,r}_{x,y}(t)]$ and $\E [A^{m,r}_{x,y}(t)]$ can be performed, which in turn allows \eqref{eq:LossStandard} and \eqref{control1} to be computed.

A MAP is a counting process with arrivals of different types governed by the transitions and holding times of another finite state Markov chain (see e.g.\ \cite{bookHe2014} and \cite{bookLatouche1999}).
Generally for MAPs each arrival is indexed by an element from a set $\mathcal C$.
For our blocking MAP we will denote the elements of this index set by the numbers 1 to $\ell$ as follows $\mathcal C_b = \{1,\dots,\ell\}$.
This notation follows from the fact that a type $k$ arrival results in $k$ lost tasks.
For our abandonment MAP the index set consists of a singleton representing an abandonment type arrival, that is $\mathcal C_a = \{a\}$.
Letting $N^{m,r}_k(t)$ be the number of type $k$ arrivals during $[0,\,t]$ when there are $m$ servers and $r$ units of buffer, it is clear from the linearity property of expectation that
\begin{equation}\label{eq:R}
\E [R^{m,r}_{x,y}(t)]  = \theta_b\sum_{k=1}^\ell k\,\E[N^{m,r}_k(t)\,|\, X(0) = x, Y(0) = y]
\end{equation}
and, letting $N^{m,r}_a(t)$ be the number of abandonments during $[0,\,t]$ when there are $m$ servers and $r$ units of buffer, we similarly have
\begin{equation}\label{eq:A}
\E [A^{m,r}_{x,y}(t)] = \theta_a\,\E[N^{m,r}_a(t)\,|\, X(0) = x, Y(0) = y]\,.
\end{equation}
Hence our focus is on determining these values.

Aside from the set $\mathcal C$, MAPs are parameterized by a sequence of matrices $(D_0, D_h, h \in \mathcal C)$ with the following properties:
\begin{itemize}
\item [(i)] the matrices $(D_h, h \in \mathcal C)$ are nonnegative;
\item [(ii)] the matrix $D_0$ has negative diagonal elements and nonnegative off diagonal elements;
\item [(iii)] the matrix $D_0$ is nonsingular; and
\item [(iv)] the matrix $D=D_0+\sum_{h \in \mathcal C} D_h$ is an irreducible infinitesimal generator.
\end{itemize}

The matrix $D$ governs the transition rates of a background Markov process, while the matrices $(D_h, h \in \mathcal C)$ specify the arrivals that are associated with relevant holding times and transitions of the background process.
The matrix $D_0$ specifies the transitions which do not have arrivals associated with them and can be calculated from property (iv).
To obtain our transient performance measures we must specify particular parameterizations of these matrices.

In our case $(X,\,Y)$ is the background process which governs the arrivals of blocked and abandoned tasks for each of our MAPs.
We encode the transitions of this background process, as given by Table~\ref{tab:rates} and illustrated for the case where arrivals only ever bring a single task in Fig.~\ref{fig:states}, in the $2(m+r+1)$ dimensional matrix $D$ with states arranged as follows:
\[
(0,1),~(1,1),~(2,1),~\dots,~(m+r,1),~(0,2),~(1,2),~\dots,~(m+r,2)\,.
\]

For our blocking MAP, when $(X,\,Y)$ is in a state $(x,\,y)$ with $x \ge m+r+2-h$ arrivals of types $k \in \{1,\dots,h\}$ occur according to Poisson processes with rates $\lambda_y^{(k)}$.
In order to record these losses we define for $y \in \{1,2\}$ the $(m+r+1)\times(m+r+1)$ square matrices $D_{y,k}$ for $k \in \{1, \dots, \ell\}$, which for diagonal elements $(i,i)$ with $i\ge m+r+2-k$ have entries $\lambda^{(k)}_y$ and 0 otherwise. For example, if $m=2$, $r=1$, and $\ell =2$ then we require
\[
D_{y,1} = \left(\begin{array}{cccc}  0& 0 & 0 & 0  \\
											0& 0 & 0 & 0  \\
											0& 0 & 0 & 0  \\
										   0 & 0& 0 &  \lambda_y^{(1)}
						   \end{array}\right)\,,\quad
D_{y,2} = \left(\begin{array}{cccc}  0& 0 & 0 & 0  \\
											0& 0 & 0 & 0  \\
											0& 0 & \lambda_y^{(2)} & 0  \\
										   0 & 0& 0 &  \lambda_y^{(2)}
						   \end{array}\right)\,,\quad
\]
for $y=1,2$.

Combining the bursty and normal arrivals (blocking losses) together we have the matrices
\begin{equation}
D_k = \left(\begin{array}{cc} D_{1,k} & 0 \\ 0 & D_{2,k} \end{array}\right)\,,\quad k = 1, \dots, \ell\,. \label{eq:Bursts}
\end{equation}

Recalling that $D_0 = D - \sum_{k=1}^\ell D_k$, based on this parameterization we are able to write down an infinite dimensional block matrix that is an infinitesimal generator of a process which describes the state of the cloud computing platform model, introduced in Section~\ref{sec:model}, augmented by a state that counts cumulative lost tasks (i.e.\ $\sum_{k=1}^\ell k\,N_k$), as follows:  
\[
Q = \left(\begin{array}{ccccccccc} D_0 & D_1 & D_2 & \cdots & D_\ell &  &  & & \\
						    & D_0 & D_1 & \cdots & & D_\ell &  & & \\
  						   & & D_0  & \cdots & &  & D_\ell & & \\
						   & &&\ddots&\ddots&\ddots&\ddots&\ddots \\
						   & &&&\ddots&\ddots&\ddots&\ddots
						   \end{array}\right)\,.
\]
In this block matrix each row of matrices corresponds to a different total number of blocked tasks during $[0,\,t]$, or the `level' of the overall blocking MAP.
The first row corresponds to no blockages, while the second row corresponds to a single blockage, and so on for the further rows.
Within each row the block $D_0$ corresponds to movements of the background or `phase' process $(X,\,Y)$, which are the transitions that are not associated with any blockages (i.e.\ of the system model).
In our case this is services, arrivals (when not at capacity), and changes between bursty and normal arrival behavior.

Similarly, for our abandonment MAP a transition of $(X,\,Y)$ from a state $(x,\,y)$ to $(x-1,\,y)$ with $x\ge m+1$ results in an arrival of type $a$ (abandonment loss) with probability
\begin{equation}\label{eq:Abandon}
(x-m)\mu_a\,\big[m\,\mu_s+(x-m)\mu_a\big]^{-1}\,,
\end{equation}
and otherwise a service has occurred. To see that this is the case, observe that when the process is in state $(x,\,y)$ with $x\ge m+1$, this implies that there must be $m$ tasks undergoing processing and $x-m$ tasks waiting in the buffer. Each of the tasks undergoing processing has an independent exponentially distributed amount of time with mean $\mu_s^{-1}$ until it completes, and each of the tasks waiting in the buffer has an independent exponentially distributed amount of time with mean $\mu_a^{-1}$ until it abandons. Equation \eqref{eq:Abandon} then follows from the properties of the exponential distribution.

To obtain an infinite dimensional block matrix that is an infinitesimal generator for $N_a$ we place the departures that correspond to an abandonment in the matrix $D_a$.
This $2(m+r+1)$ dimensional matrix has entries $i\,\mu_a$ for $i=m, m+1, \dots, m+r, 2(m+1), 2(m+2), \dots, 2(m+r)$ at coordinates $(i+1,\,i+2)$.
Now, using $D_0' = D-D_a$, we have that
\[
Q_a = \left(\begin{array}{cccc} D_0' & D_a & & \\
						    & D_0' & D_a &  \\
  						   & &\ddots & \ddots \\
						   & &&\ddots \\
						   \end{array}\right)\,.
\]
Similar to the case for $Q_k$, in the matrix $Q_a$ each row of matrices corresponds to a different total number of abandonments during $[0,\,t]$, or the level of the overall abandonment MAP.
Since only a single abandonment can occur at a time, each row can be parameterized using only $D_0'$ and $D_a$.

We have now completely defined our model of a distributed cloud computing platform and the encompassing machinery that we will use to perform a transient performance evaluation.

\section{Transient Performance Evaluation} \label{sec:performance}
The simplest application of MAMs is to find the equilibrium distribution of a finite state Markov process.
Given that the process has infinitesimal generator $Q$, if we denote the equilibrium distribution by (row vector) $\bP$ and a vector of ones with the same dimension as $\bP$ by $\bO$ then this simply amounts to solving the equation $\bP Q = 0$ subject to $\bP \bO = 1$ where each component of $\bP$ must be non-negative.
This computation is straightforward on modern computers for most finite state space Markov processes of interest.
For example, this simple computation provides an alternative method for obtaining the probability given by \eqref{eq:Erlang}.

The field of MAMs, and more broadly algorithmic probability, is concerned with augmenting computational methods, such as the one just discussed, with analytical results.
Through this it is often possible to answer questions that are computationally difficult in the absence of analysis and analytically infeasible in the absence of computational resources.
The development of the framework that we are presenting in this paper falls exactly into this category of methodology.
By utilizing the special structure of MAPs, usual Markov process theory, and our formulation of the problem, we are able to arrive at expressions that can be evaluated numerically to answer the challenging questions faced by managers of distributed cloud platforms.
In the next subsection we will derive the expected value of lost revenue during $[0,\,t]$ conditional on particular values of $(X(0),\,Y(0))$ for the unpredictable case, where the rates of the process are time homogeneous, before incorporating predictable behavior in the subsequent subsection.

\subsection{Unpredictable arrival rate expected value}

Using $(D_k : 1 \le k \le \ell)$ and $D_a$ we have shown how to construct a pair of two dimensional Markov processes, $R^{m,r}_{x,y}$ and  $A^{m,r}_{x,y}$ where the second dimension gives the value of lost tasks from blockages and abandonments respectively.
Using standard MAMs we show how to find the expected value of these processes at a finite time $t$ in the form of an analytical expression that can be evaluated numerically.
Our novel formulation of the encompassing performance evaluation model allows for systems of the type detailed in Section~\ref{sec:model} to be evaluated using Theorem~2.3.2 in \cite{bookHe2014}. 

Proposition~\ref{prop:CapVal1} (given below) summarizes the application of \cite[Theorem~2.3.2]{bookHe2014} to our illustrative model. The proposition gives the expected lost revenue for finite time horizons as a function of different choices of $m$ and $r$, the current number of tasks in the system $x$, and the current state of the unpredictable process $y$.
Through straightforward optimization of this function in terms of $m$ and $r$ at a chosen planning interval $T$ a cloud platform manager can dimension their system at the time points $0,T, 2T, \dots$ to obtain performance improvements.
We again highlight that this optimization need not {\em necessarily} be performed online, the manager only needs access to the mapping $(x,\,y) \to (m,\,r)$ for each $(x,\,y)$ of interest up to some maximal capacity value.
\begin{proposition}\label{prop:CapVal1}
For constant $(D_k : 1 \le k \le \ell)$, $D_a$, and $D$ the capacity value function can be evaluated as
\begin{align}
g_{x,y}^{m,r}(t) = (m\theta_s+r\theta_u+\bP D^* \bO)\,t -\bP_0\big(\exp(D\,t)-I\big)\,D^-D^*\bO\,, \label{eq:g}
\end{align}
where
\[
D^* = \theta_a\,D_a+\theta_b\sum_{j=1}^\ell j\, D_j\,,
\quad
D^- = (\bO\bP-D)^{-1}\,,
\]
$\bO$ is a $2(m+r+1)$ column vector of ones, $\bP$ is the equilibrium distribution of the Markov process $(X,\,Y)$, $\bP_0$ is a vector indicating that the process $(X,\,Y)$ starts in state $(x,\,y)$, and $I$ is an identity matrix of appropriate dimension.
\end{proposition}

\subsection{Incorporating predictable bursts}
If the rate at which tasks arrive to a system is constant over any particular planning horizon but varies between planning intervals, then a manager can use the framework developed in the previous subsection tailored to each interval. The purpose of this subsection is to show how our framework can be adjusted when the rate at which tasks arrive to the system is known (or believed) to change at predictable time instances {\em within} a particular planning horizon $[0,t]$.  Specifically, we adapt the framework we presented in the previous subsection to the case of a predictably time-varying arrival rate when arrival rate changes occur only at a countable number of distinct time points within a particular planning horizon.

Suppose we have $h$ distinct time periods, defined by their end points $0<t_1<t_2<\cdots<t_h:=t$, which form a partition of $[0,t]$,
each having its own arrival rate.  Denote by $D_{(i)}$, $D_{(i)}^*$, and $D^-_{(i)}$ the corresponding parameterization in time period $i$.
Then, using the standard expression for the transient distribution of a time-homogeneous finite state Markov process for each interval (for details see e.g.\ \cite[p.\ 259]{bookGS2001}),
we may compute the capacity value function as
\begin{align}
&g_{x,y}^{m,r}(t) = (m\theta_s+r\theta_u)\,t+\sum_{i=1}^h\bP_{t_{i-1}} D_{(i-1)}^* \bO(t_{i}-t_{i-1})\notag\\
& \hspace{30mm}- \sum_{i=1}^h  \bP_{0,t_{i-1}}\Big(\exp\big(D_{(i-1)}\,(t_{i}-t_{i-1})\big)-I\Big)\,D_{(i-1)}^-D_{(i-1)}^*\bO\,, \label{eq:g2}
\end{align}
where $t_0 = 0$,
\[
\bP_{0,t_i} := \bP_0\prod_{j=1}^i \exp\big(D_{(j-1)} (t_{j}-t_{j-1})\big)\,,
\]
and $\bP_{t_i}$ is the equilibrium distribution of the Markov process with infinitesimal generator $D_{(i)}$.

This expression may be used in the same way as Proposition~\ref{prop:CapVal1}. In this case the mapping $(x,y) \to (m,r)$ will also be dependent on the predictable bursts within the planning horizon.

\section{Illustrations}\label{sec:examples}
This section presents four illustrative applications of our method to models of distributed cloud computing platforms.
Each subsection illustrates a different aspect of the method.
We will first apply the method to the simplest possible setting, that of a system with only predictable and time homogeneous arrivals.
We illustrate the connection between our transient performance measure and transient decision making, as well as demonstrate that this may provide modest improvements compared to decisions based on equilibrium methods.
Subsequently, we allow arrivals to vary with time in a predictable manner and show that in this case the improvement over equilibrium methods may be quite substantial.
We then investigate unpredictable arrivals, where we again see that transient decision making can outperform equilibrium decision making, even when the decision maker is unaware of the unpredictable nature of the arrival process. Finally, we investigate the effect of batch arrivals on system performance, and again see our policy may provide improvements.

In our case the traffic intensity varies according to the value of $m$ that is selected by our provisioning framework. Hence in this section we will fix the arrival process and study the resulting changes in $m$ and $r$.  Similarly, the mean service time $\mu_s$ will be absorbed into the choice of $m$ and $r$ in the same fashion. In the final example, however, the coefficient of variation is varied through the distribution of batch size, when arrivals of this type are considered. We will evaluate the performance of our framework subject to an increasing coefficient of variation.

We note that for a cloud computing platform to be sustainable the values of the cost parameters $\theta_b$ and $\theta_s$ (see Table~\ref{tab:costs}) must be such that the expected cost of keeping a server active for the duration of a job is lower than the revenue that would be lost from a blocked job, and similarly for $\theta_a$ and $\theta_u$.

Later in this section we will give an example analysis of batchy and bursty behavior through specific parameter choices. These examples will suggest that, with our parameter choices, accounting for batchiness of the arrival process may be more important than accounting for burstiness.

\subsection{Time homogeneous $M/M/m/m{+}r$}
\begin{figure}
\centering
\includegraphics{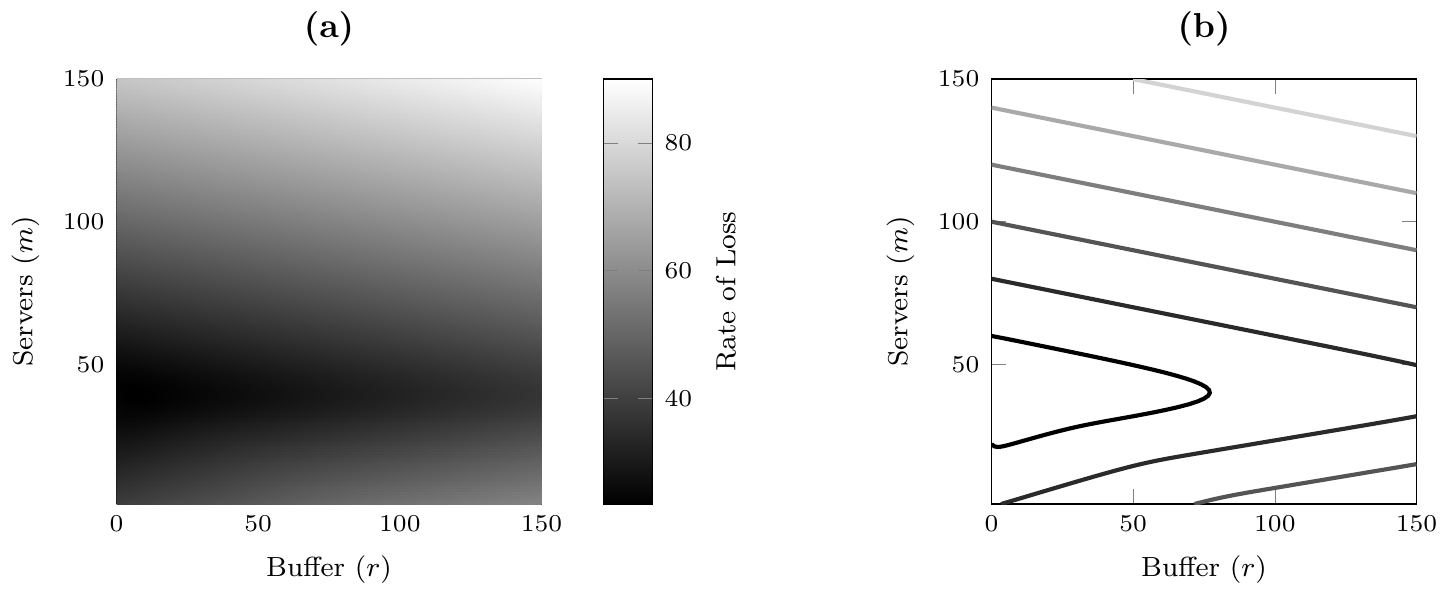}
\caption{{\bf (a)} Equilibrium rate of loss for different $m$ and $r$ combinations for equilibrium policy and {\bf (b)} contour plot of equilibrium rate of loss for different $m$ and $r$ combinations for equilibrium policy.}\label{fig:ConstantEquil}
\end{figure}

\begin{figure}
\centering
\includegraphics{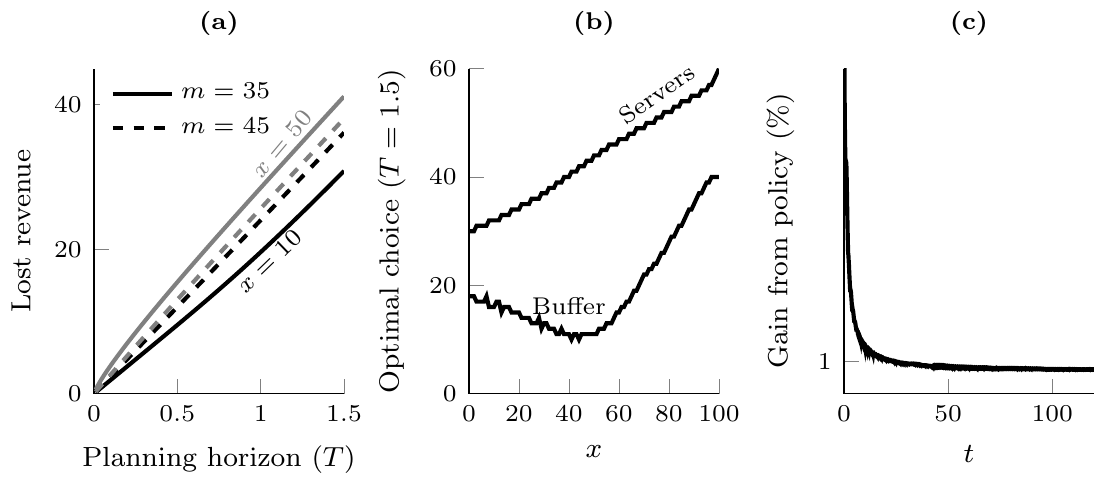}
\caption{{\bf (a)} expected losses during $[0,\,T]$ when $r=15$ for different choices of $m$ and initial condition $x$, {\bf (b)} optimal buffer ($m$) and server ($r$) choices for different initial conditions $x$, and {\bf (c)} estimated expected gain from adjusting according to the optimal choice in (b) every 1.5 time units in place of using the optimum given by (a). }\label{fig:Constant}
\end{figure}

In this subsection we will explore the simplest case of our model using Fig.~\ref{fig:ConstantEquil} and Fig.~\ref{fig:Constant}.
Assume that tasks of unit size arrive to a cloud computing platform according to a Poisson process with rate 80 (that is, we expect 80 tasks to arrive per time unit), tasks take an exponential amount of time with mean $1/2$ to be processed, and will wait in a buffer for up to an exponential time with mean $1$ to begin processing before abandoning the system.
Blockages and abandonments incur losses to the manager of sizes 0.5 and 0.55 respectively.
Units of server cost 0.5 and units of buffer cost 0.1 to operate per unit of time.
Since there are no unpredictable changes in the arrival rate we may set $\lambda_1^{(1)}=\lambda_2^{(1)}$ and arbitrarily take $\alpha=\beta=1$.

The first term in \eqref{eq:g} gives the equilibrium rate of loss from this system given a choice of $m$ and $r$.
Minimizing over this term in $(m,\,r)$ is equivalent (or superior) to many of the equilibrium performance measures considered by the papers discussed in the introduction.
In Fig.~\ref{fig:ConstantEquil} we show this function for different buffer and server choices.
It is evident that low ($<30$) or high ($>50$) values of $m$ result in a greater rate of loss (lighter shading) for a wide range of values of $r$.
For $m=40$ there is a low rate of loss (dark shading) for a wide range of values of $r$; as $r$ increases the loss rate experience a mild decrease.
In fact, if we were to only consider the equilibrium of the system in our decision making process, then we would choose to have 40 servers and 8 units of buffer; although $m$ and $r$ combinations near to this point experience a similar rate of loss.

In the first panel of Fig.~\ref{fig:Constant} we display the function \eqref{eq:g} of our system with $r=15$, lower initial tasks in the system $x=10$ (gray lines), higher initial tasks in the system $x=50$ (black lines), lower number of servers $m=35$ (solid lines), and higher number of servers $m=45$ (dashed lines).
For the lower initial tasks in the system it can be seen that choosing the lower number of servers is expected to reduce the loss incurred by the system's manager.
On the other hand, the higher number of initial tasks has reduced losses when there are more servers utilized.
This illustrates the relationship between expected losses, server and buffer choice, and the current number of tasks in the system.

Using the relationship illustrated in the first panel of Fig.~\ref{fig:Constant} we are able to choose $m$ and $r$ optimally for a planning horizon of 1.5 for each $x$, which we display in the third panel of the same figure.
It can be seen that as there are more tasks in the system a higher number of servers is optimal. Interestingly, the optimal buffer size is convex in the number of tasks, with a minimum value at the equilibrium choice.

In the third panel of Fig.~\ref{fig:Constant}, we are finally able to illustrate that the transient performance measure based framework developed in this paper provides an improvement over equilibrium based frameworks.
We compare $10^6$ sample paths of the system operating using the equilibrium server and buffer choice with $10^6$ sample paths where the server and buffer choice is adjusted each 1.5 time units according to the optimal choices displayed in the third panel. In this case our policy results in an improvement of approximately 1\% compared to fixed server and buffer sizes chosen according to equilibrium system behavior.

\subsection{$M/M/m/m{+}r$ with predictable bursts}
\begin{figure}
\centering
\includegraphics{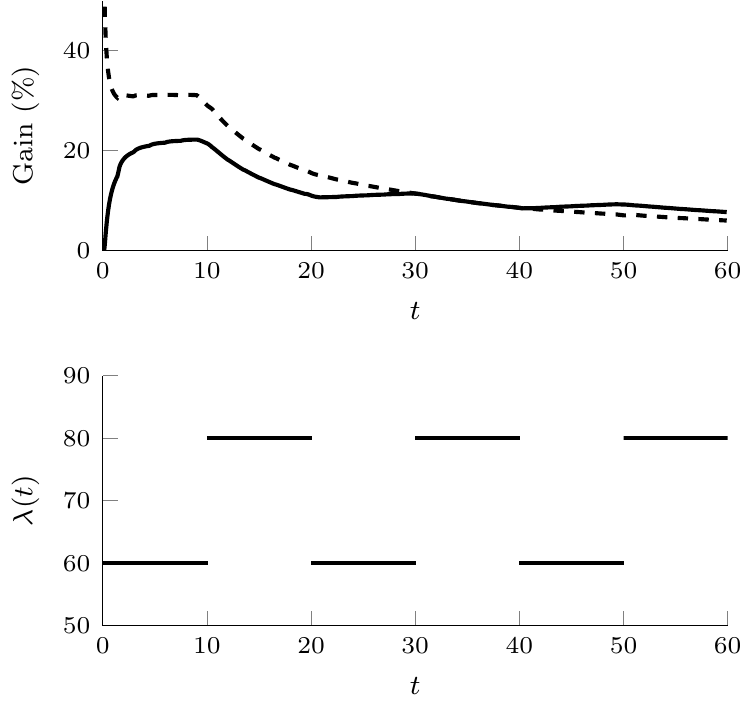}
\caption{Lower panel: predictable bursts in the arrival rate. Upper panel: gain from using our transient policy in place of a static equilibrium policy (solid) and from using our policy in place of a dynamic equilibrium policy (dashed).}
\label{fig:Pred}
\end{figure}

	This short subsection has the simple goal of highlighting that our framework may perform extremely well when a system is subject to predictable bursts in the arrival rate. Suppose that the system is the same as in the previous subsection, except that now tasks arrive at rate 60 during $\bigcup_{i\in\{0,2,4\}}[10i, 10(i+1))$ and at rate 80 otherwise, as illustrated in the lower panel of Fig.~\ref{fig:Pred}. If the system manager was to make an equilibrium based server and buffer choice, then there are three obvious alternatives: i) provision according to $\lambda = 80$ to avoid SLA violations during bursts, ii) provision according to $\lambda =60$ to avoid unnecessary operating costs during non burst periods, or iii) provision according to some mix of the previous two alternatives.
	
	The solid curve in the upper panel of Fig.~\ref{fig:Pred} compares the lost revenue of a manager who uses the first of these alternatives (over-provision with $\lambda=80$) with the lost revenue of a manager who uses our framework. This illustration suggests that the manager that adjusts server and buffer choices every 1.5 time units according to our transient provisioning framework reduces their losses by approximately 7.5\% after 60 time units, compared to the manager who provisions statically according to an equilibrium performance measure.

	The use of equilibrium performance measures does not, however, preclude the use of dynamic provisioning. The dashed curve in the upper panel of Fig.~\ref{fig:Pred} compares the lost revenue of a manager who provisions at any given time according to the equilibrium policy computed for the arrival rate of tasks at that time. In this case we still see a substantial gain from dynamically using a transient policy in place of dynamically using equilibrium policies. The strength of the transient policy likely primarily arises from its incorporation of present state information at each decision epoch (which is impossible using equilibrium methods).

\begin{figure}
\centering
\includegraphics[width=\textwidth]{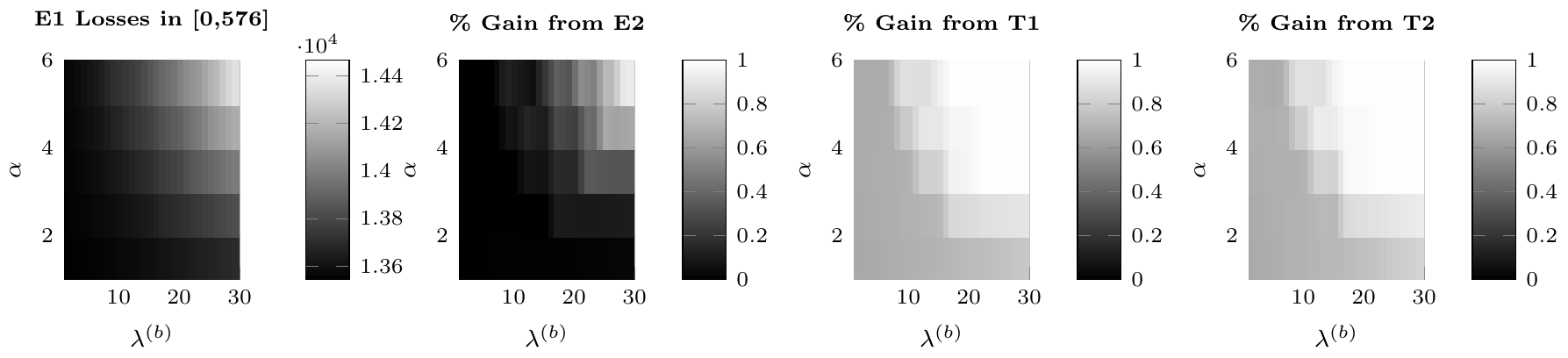}
\caption{Comparison of losses in [0, 576] for an equilibrium policy that does not account for bursty behavior (E1), an equilibrium policy that does account for bursty behavior (E2) with our transient policy when bursty behavior is not (T1) and is (T2) accounted for. }
\label{fig:Burst2}
\end{figure}

\subsection{Unpredictable time-varying $M/M/m/m{+}r$}
In this subsection we demonstrate that our framework can improve system performance in the presence of unpredictable time varying behavior.
Furthermore, we investigate and compare the effects of the burst frequency and the burst magnitude on the expected lost revenue.
The key message of the subsection, as illustrated by Fig.~\ref{fig:Burst2}, is that using our transient policy may provide an improvement on system performance, even if it is applied without knowledge that bursty behavior is occurring. We will compare the policies E1, E2, T1, and T2, where E and T correspond to equilibrium and transient decision making respectively, and 1 and 2 correspond to bursty unaware and aware decision making respectively.

\begin{figure}
\begin{center}
\begin{tikzpicture}
	\node at (0, 0) {$80-\frac{\lambda_{(b)}\alpha}{\alpha+5}$};
	\draw (0, 0) ellipse (1cm and 0.5cm);
	\node at (3.15, 0) {$80+\lambda_{(b)}$};
	\draw (3, 0) ellipse (1cm and 0.5cm);
	\draw[->, thick] (0,0.525) to  [bend right=-20] (3,0.525);
	\draw[<-, thick] (0,-0.525) to  [bend right=20] (3,-0.525);
	\node at (1.5,1) { $\alpha$};
	\node at (1.5, -1) { $5$};
\end{tikzpicture}
\end{center}
\caption{Two state background process. The left state represents the normal arrival rate while the right state represents bursty periods. The parameters $\alpha$ and $\lambda_{(b)}$ determine burst size and duration, however the average rate of arrivals is fixed at 80.}
\label{fig:Burst0}
\end{figure}
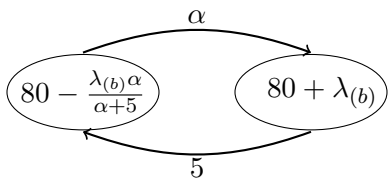

To measure burst frequency and magnitude we allow arrivals of unit size to occur according to a Markovian arrival process with two underlying states: the states correspond to a `normal' demand regime and an `increased' demand regime.
In Fig.~\ref{fig:Burst0} we depict these two states, their arrival rates, and the transition rates between them.
When the system is in the increased demand regime it moves to the normal regime at rate $5$.
In this case each time the arrival rate increases due to a burst, the increase is expected to persist for 0.2 time units.
On the other hand, when the system is in the normal regime it moves to the increased regime at rate $\alpha$.
A higher value of $\alpha$ therefore corresponds to more frequent bursts.
We assume that a burst results in the arrival rate increasing by $\lambda_{(b)}$.
Since we wish to analyze the effect of burst frequency and magnitude, in order to keep the overall time average arrival rate equal to 80 (as for the time homogenous case already considered), we counterbalance the increased demand regime arrival rate with a decrease in the normal arrival rate of $\lambda_{(b)}\alpha/(\alpha+5)$.
This follows from the fact that the ergodic distribution of the background process that alternates between normal and increased demand is $\big(5/(5+\alpha),\,\alpha/(5+\alpha)$\big).

Increases in $\lambda_{(b)}$ result in a higher arrival rate during burst periods and a decreased arrival rate during the normal regime.
Increases in $\alpha$ result in a smaller difference between normal and increased demand, but increase the frequency with which bursts occur.

Now that it is clear how we are measuring and evaluating burstiness, we return to Fig.~\ref{fig:Burst2}.
In the first panel of this figure we consider a manager who uses the equilibrium policy from the homogeneous arrival subsection --- that is, they do not account for the bursty behavior at all.
It can be seen that individual increases in $\alpha$ or $\lambda_{(b)}$ have a minor effect on losses, but that when both of these parameters increase together larger losses are incurred.

In the second panel of Fig.~\ref{fig:Burst2} the manager is aware of the values of $\alpha$ and $\lambda_{(b)}$.
Since individual increases in the burstiness parameters do not seem to affect losses substantially, we do not expect that a change in policy will be very beneficial.
This is confirmed by the second panel of Fig.~\ref{fig:Burst2}.
When the bursty parameters are high together however, we see that accounting for the presence of burstiness is beneficial, even from an equilibrium perspective.

Finally, the third panel shows that a gain of approximately 0.8--1\% is generated through the usage of our transient policy, even when the burstiness of the system is not accounted for.
Indeed, the fourth panel of the figure is highly similar to the third panel, indicating that in this example the policy performs equally well when the manager does not account for burstiness.
Notably, our transient policy performs better than the equilibrium policy, even when the equilibrium policy is bursty aware and the transient policy is bursty unaware.

\subsection{Batchiness}
After having seen that our transient policy reduces losses in a bursty system, even when the manager is unaware of the bursty behavior, we will now investigate if that is also true for a batchy system.
In this case we will define a system to be more batchy when holding the expected number of arrivals in any time interval $[0,\,t]$ to be the same, the arrivals are able to come in batches of larger size.

\begin{figure}
\centering
\includegraphics{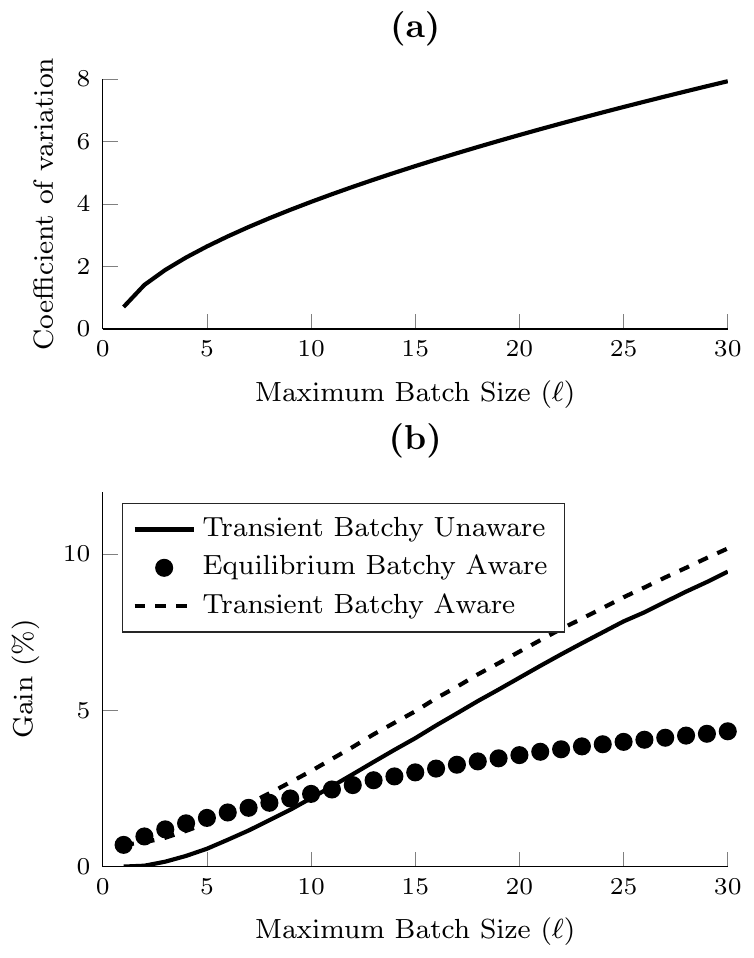}
\caption{{\bf (a)} Coefficient of variation of job size, {\bf (b)} \% reduction in lost revenue over [0, 54] from incorporating batch information into the equilibrium policy (dots), updating server and buffer sizes periodically (solid), or both incorporating batch information and using our transient policy (dashed) compared to a batchy unaware equilibrium policy over the period. }
\label{fig:Batch}
\end{figure}

To model this we fix a maximum batch size $\ell$ and let the arrival rate of batches of size $k$ be $\lambda/(k\,\ell)$.
From the basic properties of Poisson processes, we see that
\[
\sum_{k=1}^\ell \E[N_k(t)] = \sum_{k=1}^\ell k\,\frac{\lambda t}{k\,\ell} = \lambda \, t\,,
\]
so that an overall expected number of arrivals $\lambda t$ during $[0,\,t]$ is maintained.
A higher value of $\ell$ is, however, clearly more batchy. Specifically, recalling that $K$ is a random variable which governs the size of an arbitrary batch we have that
\[
{\mathbb P}(K=k) = \frac{1}{k\,H_\ell}\,,\quad k \in \{1, 2, \dots, \ell\}\,,
\]
where $H_\ell = \sum_{k=1}^\ell k^{-1}$ is a harmonic number. Since it can be shown that the coefficient of variation of a job will be
\[
\frac{\sqrt{\E[K]+\text{Var}(K)}}{\E[K]}\,,
\]
from this we can see that the coefficient of variation for an arbitrary job is
\[
\sqrt{\frac{H_l\,(\ell+2)}{2}-1}\,.
\]
This function is increasing in $\ell$ (as illustrated in the top panel of Fig.~\ref{fig:Batch}), meaning that increased batchiness (i.e.\ higher $\ell$) results in a higher coefficient of variation. In this example the coefficient of variation lies in the range of approximately 0.7071 to 7.9322.

For our ongoing illustrative parameters, an increase in batchiness (or variation) results in higher losses.
Fig.~\ref{fig:Batch} focuses on the percentage reduction in losses that can be achieved compared with provisioning according to the equilibrium policy of the earlier homogeneous section (i.e.\ $m=40$, $r=8$) during the arbitrarily chosen time period $[0,\,54]$.
The figure shows that in this case the gains from accounting for batchiness may be substantially greater than the gains from accounting for burstiness.
For lower levels of batchiness, $\ell \in \{1, \dots, 10\}$ using an equilibrium batchy aware policy appears to provide similar gains to the transient batchy aware policy, and performs better than the transient batchy unaware policy.
This contrasts with the bursty scenario where the transient policy was superior even when bursty behavior was not accounted for by the decision maker.
For higher levels of batchiness the transient policies are each superior, yielding gains of approximately 10\%, even if they are not aware that the system is batchy (with the batchy aware transient policy performing approximately 1\% better than the batchy unaware transient policy).

\section{Outlook}\label{sec:conclude}
The main contribution of this paper is the observation that the capacity value function can be found using matrix analytic methods.
Based on this observation, and upon incorporating operating costs for servers, we have presented a framework that may be used for transient provisioning and performance evaluation of cloud computing platforms. The purpose of the present paper is to introduce a framework for the analysis of models which are applicable to the domain of transient provisioning of cloud computing platforms. As such we presented our framework in the context of a model that is comparable to those which are currently state of the art in the literature of this area (e.g.\ \cite{Atmaca2016,Bruneo2014,Khazaei2011,Khazaei2012,Khazaei2013,Khazaei2013b,Maccio2015,Zhang2016}). The complexity of the illustrative model was chosen as a balance between presentation and relevance. We envisage that many possible extensions to the underlying model could be implemented within our encompassing performance evaluation framework with varying degrees of complexity.

Generalizing the system model to have more than two states in the underlying burst modulating process $Y$ follows from simply augmenting the matrix $D_k$ in \eqref{eq:Bursts} with an additional matrix $D_{y,k}$ for each additional state in $Y$. Such modifications result in a linear increase in complexity, the dimension of the matrices for which it is necessary to compute exponentials and inverses of is the number of states of $Y$ multiplied by $(m+r+1)$.  In the same way, to model a system failure that does not result in removal of tasks but simply a pause in service, the matrix $D_k$ could be augmented with a matrix representing a set of states with a reduced (or 0) service rate; which is again a linear increase in complexity.  It would also be possible to model the case that such a failure removes all (or a random number) of tasks from the system through some careful adjustment of the rates in $D$; in this case the formulation is changed but the complexity is not affected. Furthermore, it would also be possible to apply our framework in the case of multiple resource types by taking the process $X$ to be multidimensional (which is simply a notational increment in methodology).

The above additional applications of our framework rely on modifications of the structure of $g^{m, r}_{x,y}(t)$. Recall, however, that underlying our framework is the function
\[
g^{\tilde m, \tilde r}_{x,y}(t) - g^{m, r}_{x,y}(t)\,,
\]
which could also be modified. For example, by adding a term that depends on the difference $m-\tilde m$, costs related to the provisioning of new resources could be included.

Despite the flexibility of our approach, there do exist limitations to our method that remain a challenge. For example, we have assumed that abandonments occur on a task-wise basis, when in reality all of the tasks from a particular batch arrival may abandon together. Another limitation is that the model is inherently Markovian, so that we have not been able to incorporate generally distributed service times for tasks (e.g.\ to model heavy tailed behavior). It is not obvious how to extend our framework to incorporate these features.

Another avenue of future research could be aimed at obtaining higher order moments of
\[
R^{m,r}_{x,y}(t)+A^{m,r}_{x,y}(t)+M^{m,r}_{x,y}(t)\,.
\]
Doing so would enable risk taking preferences to be incorporated into the decision making process that we have developed and allow more sophisticated performance evaluation.

{\footnotesize \section*{Acknowledgements}
The authors are grateful to the referees for providing valuable feedback that greatly improved the content of the paper, in particular the section on predictable bursts benefited substantially from this feedback. The authors thank Yoni Nazarathy (The University of Queensland) for discussions that were crucial to the development of the framework presented in this paper. The authors are also grateful to Michel Mandjes (University of Amsterdam) for detailed comments on a draft manuscript. BP acknowledges the support of  an Australian Government Research Training Program (RTP) Scholarship, the Australian Research Council (ARC) through the ARC Centre of Excellence for the Mathematical and Statistical Frontiers (ACEMS) under grant number CE140100049, and the support of the Netherlands Organisation for Scientific Research (NWO) through the Gravitation project N{\sc etworks} under grant number 024.002.003.

\bibliographystyle{plain}
{\footnotesize
\bibliography{CapPricing}}

\begin{thebibliography}{10}

\bibitem{QueueBook}
Queueing networks.
\newblock volume 154 of {\em International Series in Operations Research \&
  Management Science}. 2011.

\bibitem{AmazonSite}
Amazon auto-scaling.
\newblock \url{https://aws.amazon.com/autoscaling}, 2017.

\bibitem{Abate1998}
J.~Abate and J.~Whitt.
\newblock Calculating transient characteristics of the {E}rlang loss model by
  numerical transform inversion.
\newblock {\em Communications in statistics. Stochastic Models.},
  14(3):663--680, 1998.

\bibitem{Atmaca2016}
T.~Atmaca, T.~Begin, A.~Brandwajn, and H.~Castel-Taleb.
\newblock Performance evaluation of cloud computing centers with general
  arrivals and service.
\newblock {\em Parallel and Distributed Systems, IEEE Transactions on}, 2015.

\bibitem{Braunsteins2016}
P.~Braunsteins, S.~Hautphenne, and P.~G. Taylor.
\newblock The roles of coupling and the deviation matrix in determining the
  value of capacity in ${M}/{M}/1/{C}$ queues.
\newblock {\em Queueing Systems}, 83(1):157--179, 2016.

\bibitem{Breuer2002}
L.~Breuer.
\newblock An {EM} algorithm for batch {M}arkovian arrival processes and its
  comparison to a simpler estimation procedure.
\newblock {\em Annals of Operations Research}, 112:123--138, 2002.

\bibitem{Bruneo2014}
Dario Bruneo.
\newblock A stochastic model to investigate data center performance and qos in
  iaas cloud computing systems.
\newblock {\em Parallel and Distributed Systems, IEEE Transactions on},
  25(3):560--569, 2014.

\bibitem{Buchholz2010}
P.~Buchholz, P.~Kemper, and J.~Kriege.
\newblock Multi-class {M}arkovian arrival processes and their parameter
  fitting.
\newblock {\em Performance Evaluation}, 67:1092--1106, 2010.

\bibitem{Buchholz2009}
P.~Buchholz and J.~Kriege.
\newblock A heuristic approach for fitting {MAP}s to moments and joint moments.
\newblock In {\em Quantitative Evaluation of Systems, Sixth International
  Conference on}, pages 53--62. IEEE Computer Society, 2009.

\bibitem{Buchholz2017}
Peter Buchholz and Jan Kriege.
\newblock Fitting correlated arrival and service times and related queueing
  performance.
\newblock {\em Queueing Systems}, 2017.

\bibitem{Casale2008}
Giuliano Casale, Eddy~Z Zhang, and Evgenia Smirni.
\newblock Kpc-toolbox: Simple yet effective trace fitting using {M}arkovian
  arrival processes.
\newblock In {\em Quantitative Evaluation of Systems, 2008. QEST'08. Fifth
  International Conference on}, pages 83--92. IEEE, 2008.

\bibitem{Chiera2005}
B.~A. Chiera, A.~E. Krzesinski, and P.~G. Taylor.
\newblock Some properties of the capacity value function.
\newblock {\em SIAM Journal on Applied Mathematics}, 65(4):1407--1419, 2005.

\bibitem{Chiera2002}
B.~A. Chiera and P.~G. Taylor.
\newblock What is a unit of capacity worth?
\newblock {\em Probability in the Engineering and Informational Sciences},
  16(4):513--522, 2002.

\bibitem{Erlang1918}
A.~K. Erlang.
\newblock Solutions of some problems in the theory of probabilities of
  significance in automatic telephone exchanges.
\newblock {\em The Post Office Electrical Engineers' Journal}, 10:189--197,
  1918.

\bibitem{Ghosh2010}
Rahul Ghosh, Francesco Longo, Vijay~K Naik, and Kishor~S Trivedi.
\newblock Quantifying resiliency of iaas cloud.
\newblock In {\em Reliable Distributed Systems, 2010 29th IEEE Symposium on},
  pages 343--347. IEEE, 2010.

\bibitem{bookGS2001}
G.~Grimmett and D.~Stirzaker.
\newblock {\em Probability and Random Processes}.
\newblock Oxford University Press, 2001.

\bibitem{bookHarcholBalter2013}
M.~Harchol-Balter.
\newblock {\em Performance Modeling and Design of Computer Systems: Queueing
  Theory in Action}.
\newblock Cambridge University Press, 2013.

\bibitem{bookHe2014}
Q.~He.
\newblock {\em Fundamentals of Matrix-Analytic Methods}.
\newblock Springer, 2014.

\bibitem{Kelly1991}
F.~P. Kelly.
\newblock Loss networks.
\newblock {\em The Annals of Applied Probability}, 1(3):319--378, 1991.

\bibitem{Khazaei2011}
Hamzeh Khazaei, Jelena Mi{\v{s}}i{\'c}, and Vojislav~B Mi{\v{s}}i{\'c}.
\newblock Performance analysis of cloud centers under burst arrivals and total
  rejection policy.
\newblock In {\em Global Telecommunications Conference (GLOBECOM 2011), 2011
  IEEE}, pages 1--6. IEEE, 2011.

\bibitem{Khazaei2012}
Hamzeh Khazaei, Jelena Mi{\v{s}}i{\'c}, and Vojislav~B Mi{\v{s}}i{\'c}.
\newblock Performance analysis of cloud computing centers using {M}/{G}/m/m{+}r
  queuing systems.
\newblock {\em Parallel and Distributed Systems, IEEE Transactions on},
  23(5):936--943, 2012.

\bibitem{Khazaei2013}
Hamzeh Khazaei, Jelena Mi{\v{s}}i{\'c}, and Vojislav~B Mi{\v{s}}i{\'c}.
\newblock Performance of cloud centers with high degree of virtualization under
  batch task arrivals.
\newblock {\em IEEE Transactions on Parallel and Distributed Systems},
  24(12):2429--2438, 2013.

\bibitem{Khazaei2013b}
Hamzeh Khazaei, Jelena Mi{\v{s}}i{\'c}, Vojislav~B Mi{\v{s}}i{\'c}, and Saeed
  Rashwand.
\newblock Analysis of a pool management scheme for cloud computing centers.
\newblock {\em IEEE Transactions on parallel and distributed systems},
  24(5):849--861, 2013.

\bibitem{Knessl2015}
C.~Knessl and J.~S.~H. van Leeuwaarden.
\newblock Transient analysis of the {E}rlang {A} model.
\newblock {\em Mathematical Methods of Operations Research}, 82:143--173, 2015.

\bibitem{bookLatouche1999}
Guy Latouche and Vaidyanathan Ramaswami.
\newblock {\em Introduction to matrix analytic methods in stochastic modeling},
  volume~5.
\newblock Siam, 1999.

\bibitem{Maccio2015}
Vincent~J Maccio and Douglas~G Down.
\newblock On optimal policies for energy-aware servers.
\newblock {\em Performance Evaluation}, 90:36--52, 2015.

\bibitem{Mandjes2001}
M.~Mandjes and A.~Ridder.
\newblock A large deviations approach to the transient of the {E}rlang loss
  model.
\newblock {\em Performance Evaluation}, 43:181--198, 2015.

\bibitem{Mann2015}
Zolt{\'a}n~{\'A}d{\'a}m Mann.
\newblock Allocation of virtual machines in cloud data centers --- a survey of
  problem models and optimization algorithms.
\newblock {\em ACM Computing Surveys}, 48(1):11, 2015.

\bibitem{Moler1978}
Cleve Moler and Charles Van~Loan.
\newblock Nineteen dubious ways to compute the exponential of a matrix.
\newblock {\em SIAM Review}, 20(4):801--836, 1978.

\bibitem{Patch2015}
Brendan Patch, Thomas Taimre, and Yoni Nazarathy.
\newblock Performance of faulty loss systems with persistent connections.
\newblock {\em ACM SIGMETRICS Performance Evaluation Review}, 43(2):16--18,
  2015.

\bibitem{Sidje1998}
Roger~B. Sidje.
\newblock Expokit: A software package for computing matrix exponentials.
\newblock {\em ACM Trans. Math. Softw.}, 24(1):130--156, March 1998.

\bibitem{Tan2015}
Yue Tan and Cathy~H Xia.
\newblock An adaptive learning approach for efficient resource provisioning in
  cloud services.
\newblock {\em ACM Sigmetrics Performance Evaluation Review}, 42(4):3--11,
  2015.

\bibitem{Leeuwaarden2016}
Johan van Leeuwaarden, Britt Mathijsen, and Fiona Sloothaak.
\newblock Cloud provisioning in the {QED} regime.
\newblock In {\em Proceedings of the 9th EAI International Conference on
  Performance Evaluation Methodologies and Tools}, pages 180--187. ICST
  (Institute for Computer Sciences, Social-Informatics and Telecommunications
  Engineering), 2016.

\bibitem{Zhang2012}
Qi~Zhang, Mohamed~Faten Zhani, Shuo Zhang, Quanyan Zhu, Raouf Boutaba, and
  Joseph~L Hellerstein.
\newblock Dynamic energy-aware capacity provisioning for cloud computing
  environments.
\newblock In {\em Proceedings of the 9th International Conference on Autonomic
  Computing}, pages 145--154. ACM, 2012.

\bibitem{Zhang2016}
Sheng Zhang, Zhuzhong Qian, Zhaoyi Luo, Jie Wu, and Sanglu Lu.
\newblock Burstiness-aware resource reservation for server consolidation in
  computing clouds.
\newblock {\em IEEE Transactions on Parallel and Distributed Systems},
  27(4):964--977, 2016.

\end{thebibliography}
\end{document}